\PassOptionsToPackage{unicode}{hyperref}
\PassOptionsToPackage{hyphens}{url}
\documentclass[
]{article}
\usepackage{xcolor}
\usepackage{amsmath,amssymb}
\usepackage{iftex}
\ifPDFTeX
  \usepackage[T1]{fontenc}
  \usepackage[utf8]{inputenc}
  \usepackage{textcomp} 
\else 
  \usepackage{unicode-math} 
  \defaultfontfeatures{Scale=MatchLowercase}
  \defaultfontfeatures[\rmfamily]{Ligatures=TeX,Scale=1}
\fi
\usepackage{lmodern}
\ifPDFTeX\else
\fi
\IfFileExists{upquote.sty}{\usepackage{upquote}}{}
\IfFileExists{microtype.sty}{
  \usepackage[]{microtype}
  \UseMicrotypeSet[protrusion]{basicmath} 
}{}
\makeatletter
\@ifundefined{KOMAClassName}{
  \IfFileExists{parskip.sty}{%
    \usepackage{parskip}
  }{
    \setlength{\parindent}{0pt}
    \setlength{\parskip}{6pt plus 2pt minus 1pt}}
}{
  \KOMAoptions{parskip=half}}
\makeatother
\usepackage{graphicx}
\makeatletter
\newsavebox\pandoc@box
\newcommand*\pandocbounded[1]{
  \sbox\pandoc@box{#1}%
  \Gscale@div\@tempa{\textheight}{\dimexpr\ht\pandoc@box+\dp\pandoc@box\relax}%
  \Gscale@div\@tempb{\linewidth}{\wd\pandoc@box}%
  \ifdim\@tempb\p@<\@tempa\p@\let\@tempa\@tempb\fi
  \ifdim\@tempa\p@<\p@\scalebox{\@tempa}{\usebox\pandoc@box}%
  \else\usebox{\pandoc@box}%
  \fi%
}
\def\fps@figure{htbp}
\makeatother
\usepackage{bookmark}
\IfFileExists{xurl.sty}{\usepackage{xurl}}{} 
\hypersetup{
  hidelinks,
  pdfcreator={LaTeX via pandoc}}

\title{How online misinformation works: a costly signalling perspective}
\author{Neri Marsili\thanks{Universidad Nacional de Educación a Distancia (UNED), Madrid, Spain. Email: neri@fsof.uned.es}}
\date{\today}

\begin{document}

\maketitle
\begin{abstract}This chapter explores how online communication, particularly on
social media, reshapes the reputational incentives that motivate
speakers to communicate truthfully. Drawing on costly signalling theory
(CST), it examines how online contexts alter the social mechanisms that
sustain honest communication. Key characteristics of online spaces are
identified and discussed, namely (i) the presence of novel speech acts
like reposting, (ii) the gamification of communication, (iii)
information overload, (iv) the presence of anonymous and unaccountable
sources and (v) the increased reach and persistence of online
communication. Both epistemic pitfalls and potential benefits of these
features are discussed, identifying promising avenues for further
empirical investigation, and underscoring CST\textquotesingle s value
for understanding and tackling online misinformation.\end{abstract}

\subsection{Studying misinformation
online}\label{studying-misinformation-online}

Misinformation\footnote{I'm here adopting standard terminology,
  according to which \emph{misinformation,} unlike
  \emph{disinformation,} need not be intended to communicate falsehoods
  (see also Harris 2023a).} predates the digital era. From ancient myths
and medieval superstitions to modern urban legends, false beliefs have
circulated throughout history, spreading both online and offline.
According to a common narrative (prevalent in public discourse and
mainstream media), however, with digital technologies we have seen an
unprecedented increase in misinformation -- a radical surge for which
public commentators have coined various apocalyptic labels, such as
`infodemic' (Simon and Camargo 2023) or `the post-truth era' (McIntyre
2018).

Meanwhile, scientists offered more measured diagnoses of the phenomenon
(Scheufele, Krause, and Freiling 2021). Evidence suggests that digital
technologies have equally facilitated the spread of \emph{accurate}
information (Acerbi 2019, chap. 6; Acerbi, Altay, and Mercier 2022),
that people are generally adept at distinguishing fake news from genuine
news (Schulz, Fletcher, and Popescu 2020; Pfänder and Altay 2023), that
only a small minority of internet users is responsible for consuming and
sharing the majority of misinformation that circulates online (Grinberg
et al. 2019; Yang et al. 2021; Budak et al. 2024), and that
it\textquotesingle s unclear that exposure to misinformation causes the
acquisition of genuine false beliefs, rather than (e.g.) the other way
around (Allcott and Gentzkow 2017; Kahan et al. 2017; Enders et al.
2023; Zilinsky et al. 2024)\footnote{These findings are simplified for
  brevity; readers are encouraged to consult the original sources for
  detailed discussion.}.

Be that as it may, it\textquotesingle s important not to underestimate
the phenomenon. Even if fake news only represents a relatively small
fraction of our news diet (estimates fall between 1\% and 19.4\% -- see
Altay, Nielsen, and Fletcher 2022), this proportion is still alarming,
especially when one considers that these figures persist despite
extensive efforts to curb misinformation. Additionally, the
\emph{perception} that misinformation is prevalent online contributes to
a decline in trust in the media and a reduced willingness to engage with
credible news sources\footnote{Acerbi, Altay, and Mercier 2022; Mitchell
  2020; Fedeli 2019; Pfänder and Altay 2023; Newman et al. 2023, 25.},
which constitutes an equally alarming tendency (Rini 2021; Harris
2024a). Even as we abandon the `apocalyptic' narrative common in public
discourse, we should be keenly aware that digital technologies present
unique challenges to the integrity of our information ecosystem.

Much research on these threats has focused on the consumption of online
misinformation: how and why we end up believing misinformation. This
research typically highlights certain relevant cognitive vulnerabilities
(e.g. confirmation bias, myside bias, etc., cf. Pennycook and Rand 2021)
and the psychological mechanisms that favour misinformation uptake.
Perhaps comparatively less research has been concerned with the
production and dissemination of misinformation: the social forces that
facilitate its spread, and those that can curb it. Such is the focus of
this chapter, which explores how online environments reshape the
reputational mechanisms that sustain honest communication.

Methodologically, this chapter takes inspiration from costly-signalling
theory (CST), an approach that emerged from evolutionary biology. CST
emphasises the role played by reputational and social costs in keeping
communication truthful and reliable. From this perspective,
understanding online misinformation requires studying how digital
communication technologies alter the reputational incentives that
motivate us to be truthful. Although CST is occasionally invoked in
discussions of online misinformation (e.g. Altay, Hacquin, and Mercier
2020; Reimann 2022; Acemoglu, Ozdaglar, and Siderius 2024), there
remains no systematic, foundational framework for investigating how
digital environments (especially social media) alter the reputational
incentives that keep misinformation at bay. This chapter aims to fill
this gap. After introducing a novel CST account of how social policing
underpins reliable communication (§1--2), it examines how digital
platforms alter this fragile infrastructure of costs and rewards
(§3--7). The ultimate goal is to understand how digital technologies
alter the social mechanisms that shape our communicative exchanges,
enabling us to delineate better solutions (policies, platform design
strategies, etc.) for containing online misinformation.\footnote{For
  some previous attempt to analyse online communication with the tools
  of CST, see Donath 2007, Bergamaschi Ganapini 2021, and Reimann 2022.}

\subsection{Costly signalling theory}\label{costly-signalling-theory}

Why do communicators refrain from lying, even when being sincere goes
against their immediate interest? The central idea behind costly
signalling theory is that there are costs attached to sending deceptive
signals, and that these costs motivate senders to be truthful and
reliable

A classic illustration of a costly signal is the peacock's vibrant tail.
The peacock's tail signals its fitness to potential mates. The more
conspicuous the tail, the more attractive the peacock is to peahens.
Maintaining a huge and conspicuous tail, however, is metabolically
expensive, and it increases risk of predation. Consequently, growing a
huge tail is a signal that only a genuinely fit peacock can afford. Here
costs explain why the signal is reliable, despite the incentive to
deceive (Zahavi and Zahavi 1997; Grafen 1990).

Costs need not be linked to the \emph{production} of a dishonest signal,
as with the peacock's tail.\footnote{Costs that have this feature are
  known as \emph{handicaps}.} Some costs are rather linked to the
\emph{consequences} of sending a dishonest signal. Take the example of
`status badges' on male passerines (\emph{Passer Domesticus, Zonotrichia
querula}). Status badges signal social dominance and the ability to win
fights. This allows passerines to resolve conflicts without engaging in
potentially harmful fights (Krebs and Dawkins 1984; Smith and Harper
2003). It has been observed that when a sparrow's badge is identified as
misleading, some form of punishment (such as harassment, fighting, or
exclusion) occurs. This form of retaliation, in turn, is thought to
prevent deceptive badges from evolving (Møller 1987; Tibbetts and Dale
2004; Tibbetts and Izzo 2010). Here the proliferation of deception is
avoided by a `deterrence mechanism': a socially inflicted cost that
makes deception an unpalatable strategy.\footnote{To fully deserve the
  label of \emph{deterrent,} the costs must be high enough to offset the
  potential benefits of deception.}

Direct retaliation (also known \emph{partner-control)} isn't the only
way to socially deter deception. Social costs can also be imposed
through \emph{partner-choice} (or \emph{reputation management}). Several
species (from ground squirrels to vervet monkeys) have been observed to
selectively ignore or ostracise individuals who frequently emit
unreliable signals (Kneer and Marsili 2025, sec. 2.1). This imposes
costs on dishonest signallers, in terms of diminished influence and
social standing. The looming threat of these costs, in turn, works as a
\emph{deterrent} against dishonest communication, incentivising senders
to communicate truthfully.

In human communication, too, \emph{reputational} and \emph{social costs}
constitute key incentives to communicate truthfully and reliably
(Mercier 2020; Reid et al. 2020), and play an important role in
containing misinformation. If this is right, the theoretical tools
developed within CST can be deployed to better understand how
reputational incentives shape truthful communication and influence
misinformation spread, both offline and online.

\subsection{Costly signals in human
communication}\label{costly-signals-in-human-communication}

Human signalling systems differ significantly from those of non-human
animals. Human languages are more complex, evolved for different
purposes, and can take advantage of sophisticated cognitive faculties
and increasingly efficient means of cultural transmission.

When reputation is discussed in relation to human communication, it is
often emphasised that \emph{reputational costs} are accompanied by
equally important \emph{reputational rewards}. We keep track not only of
deceivers, but also of who is helpful and reliable. Just as spreading
false rumours can \emph{damage} your reputation, then, sharing accurate
information can \emph{improve} it. Reliable informers can gain status
and prestige, improving their influence within a community (Baumeister,
Zhang, and Vohs 2004). Henceforth, I will talk of \emph{reputational
incentives} to refer to both of these dimensions (costs and rewards)
that motivate communicators to share reliable testimony.

Crucially, humans don't need to acquire information about each other's
reputation through direct observation. We share reputational information
through `gossiping' -- a technical term for conversations about an
absent third party (Foster 2004; Dores Cruz et al. 2021). Gossip is
prevalent in human societies (Besnier 2019; Robbins and Karan 2020) and
is thought to play an important social role in spreading reputational
information (both positive and negative), especially in societies where
direct observation is difficult (Tan, Jiang, and Ma 2023). Gossiping is
much more effective than direct observation; it allows information about
reputation to circulate quickly and widely. As such, it creates much
stronger incentives to communicate truthfully, since every individual
interaction could potentially affect one's reputation at a collective
level (both positively and negatively).

Although my focus is primarily on the \emph{epistemic} dimension of
communicative costs, it's worth stressing that a message can be costly
for many other reasons: it may be offensive, unpopular, hostile,
impolite, inappropriate, immoral, etc. (Brown and Levinson 1987; McGowan
2019; Saul 2024). Non-epistemic costs and rewards play an equally
central role in motivating speakers\footnote{For example, the
  \emph{social} \emph{rewards} of signalling agreement with a desired
  group can outweigh the \emph{epistemic costs} of being insincere about
  one's actual preferences -- as studied in the literature on
  ``preference falsification'' (Frank 1996; Brady, Crockett, and Van
  Bavel 2020). As a further complication, sometimes speakers are
  motivated to send ostensibly false signals, which carry high epistemic
  costs, to reliably signal ingroup commitment (yielding high social
  rewards) -- as noted by scholars who discuss the endorsement of
  apparently irrational religious (Irons 2001) and conspiratorial
  (Bergamaschi Ganapini 2021; Williams 2022) beliefs.} and in sustaining
cooperative communication. To keep discussion manageable, however, this
chapter will restrict its focus to the \emph{epistemic} dimension of
reputational incentives\footnote{Accordingly, I'll use `reputation' as a
  shorthand for `epistemic reputation', unless otherwise specified. By
  abstracting away from non-epistemic incentives, I do not mean to
  suggest that they are of secondary importance. My goal is rather to
  lay foundations for researching online speech within a CST framework,
  setting the stage for a more comprehensive analysis in future work.},
and their impact on online misinformation.

What makes communicative behaviour epistemically appropriate? Humans
distinguish \emph{lies} from \emph{honest mistakes}, attaching different
reputational costs to each (Kneer 2018). According to an influential
view in linguistics and philosophy, this is because speech acts are
governed by \emph{epistemic norms} (Grice 1989; Searle 1969; Williamson
1996). Scholars disagree about the exact content of such norms (Pagin
and Marsili 2021, sec. 5.1). However, recent empirical evidence suggests
that assertions are considered epistemically permissible when they are
sincere and supported by appropriate evidence (Kneer 2018; 2021; Reuter
and Brössel 2019; Marsili and Wiegmann 2021).\footnote{This goes against
  some initial findings, now regarded as problematic, that suggested
  that the norm of assertion is knowledge (for an overview, see Graham
  and Pedersen 2024; Kneer and Marsili 2025). Although the focus here is
  on assertions, other speech acts (such as promises or predictions)
  generate epistemic expectations as well (Alston 2000; Watson 2004;
  Marsili 2016).} Speakers incur reputational costs when they violate
these epistemic norms (or commitments, cf. Green 2009; 2023; Graham
2020; Bruner 2024). Receivers thus distinguish between falsehoods that
are genuine infractions (e.g. lies), those that aren't (e.g. ironic
statements, hyperbole, etc.), and those that are sub-optimal but
permissible (e.g. honest mistakes).

What about reputational \emph{rewards}? Some authors suggest that
communication is governed by epistemic goals, too -- for instance,
assertions are said to be successful when they are true (Grice 1989;
Stalnaker 1978; Roberts 2012; Marsili 2018; 2023b) and relevant (Sperber
and Wilson 1995; Wilson and Sperber 2002). Presumably, speakers can
improve their epistemic reputation by reliably meeting such goals -- not
their individual conversational goals (e.g. persuading), but the mutual
goals that one accepts when one enters a conversation (e.g. sharing
information that is relevant and accurate).

To slightly complicate matters, human languages are rich in expressions
designed to signal the speaker's degree of confidence, allowing us to
fine-tune our testimony to the quality of the evidence available.
Scholars have noted an interesting trade-off between persuasiveness and
reputation here. Confident statements (such as `Certainly, \emph{p'})
pair high risk with high rewards: they are more persuasive, and affect
reputation more (for better and for worse). Hedged statements (such as
`Maybe, p') have the opposite profile: they are less persuasive and
affect reputation less. In short, speakers aren't only expected to
communicate truthfully: they are expected to calibrate their degree of
confidence to the evidence available to them (Tenney, Spellman, and
MacCoun 2008; Pozzi and Mazzarella 2023).

Of course, violations need to be detected before they can affect
reputation. Some scholars argue that humans have evolved a set of
cognitive abilities designed precisely to keep dishonest communicators
in check (\emph{epistemic vigilance,} or
\emph{monitoring}\footnote{Epistemologists often draw a distinction
  between `active monitoring' and `counterfactual sensitivity to
  defeaters': roughly, either \emph{distrusting} a speaker unless one
  has reasons to, or \emph{trusting} a speaker unless one has reasons
  not to. My understanding of `monitoring' cuts across this distinction:
  the latter, more tolerant attitude also qualifies as monitoring (cf.
  Goldberg and Henderson 2006).}\emph{)}. Rather than gullibly accepting
testimony, we are naturally led to critically evaluate the plausibility
of a piece of testimony\footnote{Though still debated, this view is
  supported by developmental evidence (Mascaro and Sperber 2009; Isella,
  Kanngiesser, and Tomasello 2019).} and to distinguish between
trustworthy and untrustworthy sources (Sperber et al. 2010). Although
epistemic vigilance is fallible (we don't detect all lies and mistakes),
it's effective enough for people to collectively detect, in the long
term, a good proportion of epistemic infractions. This, in turn, ensures
that speakers have an incentive to be truthful (Goldberg 2011; Sperber
2013; Grodniewicz 2022), as illustrated in Figure 1.

\includegraphics[width=4.97297in,height=2.33222in]{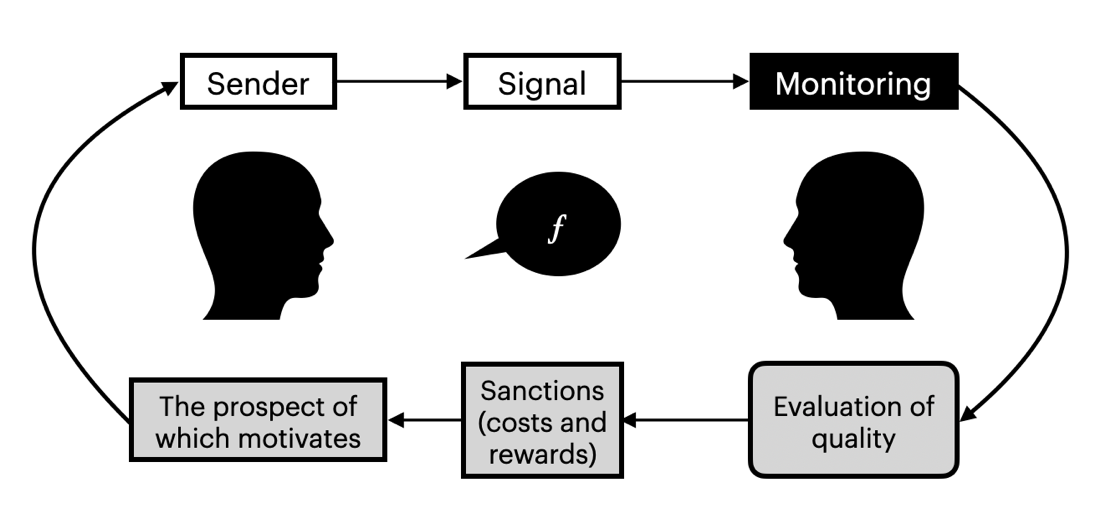}

\emph{Figure 1: A simplified representation of the model described so
far, visualising how the prospect of costs and rewards (brought about by
collective monitoring) motivates truthful communication.}

Finally, there is a peculiarity of human communication that has been
overlooked in previous work extending CST to humans. Humans have
developed strategies to minimise the chances that they will incur
reputational losses when they are caught violating a norm. Specifically,
speakers can mitigate the risk of incurring reputational costs by
strategically seeking `deniability'. Lee and Pinker (2010, 785) offer
the example of a carefully phrased bribe offer: `Gee, officer. I was
thinking that maybe the best thing would be to take care of the ticket
here, without going through a lot of paperwork'.

Strategic deniability preserves the possibility of disavowing any
intention to convey a problematic message (in this case, that one is
offering to bribe the officer). Its hallmark is the possibility of
consistently (though often insincerely) clarifying: `That's not what I
meant; you misunderstood me'. Deniability is frequently sought in
communicative interactions that are socially risky, such as romantic
advances, threats, verbal deception, and political speech.\footnote{Among
  the forms of strategic deniability that have recently attracted
  scholarly interest, it's worth mentioning \emph{insinuation} (Camp
  2018; Oswald 2022), \emph{dogwhistles} (Saul 2018; 2024),
  \emph{misleading} \emph{speech} (Adler 1997; Saul 2012; Viebahn 2021;
  Marsili and Löhr 2022), and \emph{figleaves} (Saul 2017; 2024), Often,
  these phenomena are investigated in relation to \emph{hate speech}
  (Domínguez-Armas and Soria-Ruiz 2021; Domínguez-Armas, Soria-Ruiz, and
  Lewiński 2023; Saul 2024) and other forms of \emph{harmful discourse}.}
The goal is to mitigate the (risks of paying the) costs associated with
communication. This significantly complicates the account advanced by
traditional CST, because it shows that speakers also have ways to send
uncooperative messages while minimising the risks of paying the
associated costs. This insight will prove particularly important to
understanding online communication, where technological innovations open
new opportunities for dishonest signallers to exploit communicative
deniability.

In what follows, I will apply this model to better understand the
normative profile of social media platforms, shedding light on
underexplored factors that contribute to the spread of misinformation
online. While previous research has focused on singling out isolated
factors, this chapter will aim for a more systematic explanation that
brings them together under a unified explanatory project. It will offer
a philosophical exploration of how digital platforms reshape our
epistemic environments, in the spirit of system-oriented social
epistemology (Goldman 2010).

\subsection{New communicative acts}\label{new-communicative-acts}

To study how online communication reshapes reputational incentives, a
natural first step is to analyse the new signals and conversational
contexts that are distinctive of online communication. It will be
helpful here to zoom in on social media (i.e. social networking sites),
which stand out precisely because they introduce new communicative
contexts and signals: they create, so to say, new `communicative
affordances'.

Many speech acts that we perform online (such as \emph{reposting,
liking,} or \emph{tagging}) have no direct counterpart in offline
communication. Among these novel speech acts, reposting deserves
particular attention. Reposting is a new way in which information
circulates online. Novel as they are, reposts are somewhat ambiguous in
their communicative profile. X (formerly Twitter) users often claim that
`a retweet is not an endorsement' -- meaning that a repost doesn't
necessarily express agreement with the target post. However, it's hard
to pin down which exact attitude is communicated by reposting, and
substantial philosophical\footnote{Rini 2017; Arielli 2018; Marsili
  2021; Scott 2021; 2022; Pepp, Michaelson, and Sterken 2019;
  Bergamaschi Ganapini 2021; Frigerio and Tenchini 2023.} and
empirical\footnote{boyd et al.2010; Metaxas et al. 2015; Metaxas and
  TTRT 2017; Majmundar et al. 2018; Marsili et al. 2025.} work has
attempted to address precisely this question. Interestingly, scholars
converge on noting that (partly due to this ambiguity) reposts afford
the speaker some degree of plausible deniability.

The online behaviour of the current US President Donald Trump
illustrates this egregiously. Trump has frequently retweeted conspiracy
theories and blatantly false claims. Confronted publicly about this,
he's always brushed off accusations with confidence. For instance,
challenged to explain why he retweeted a message from a neo-Nazi account
(containing made-up crime statistics about Afro-Americans), he replied:
`Bill, am I gonna check every statistic? All it was is a retweet. It
wasn't from me' (Mercieca 2016).\footnote{The example illustrates how
  reposters retain deniability. While in this specific instance one may
  find Trump's denial \emph{implausible}, deniability (as anticipated
  earlier) is a matter of being able to \emph{consistently} (or, at the
  very least, \emph{felicitously,} cf. Marsili 2023a, 1036--37; 2024,
  sec. IV) deny that an available interpretation of the utterance was
  intended.}

More generally, reposting affords the speaker some degree of
deniability, reducing reputational risks for dishonest communicators
(Marsili et al. 2025). From a CST perspective, this means that reposting
represents a convenient communicative vehicle for misinformation. It
carries many of the same benefits of explicit posting: it can pass a
message along, persuading recipients to think or act as if the content
were true. At the same time, it keeps open the option of disowning
authorship and responsibility, should the content turn out to be false,
problematic, or otherwise undesirable (cf. Mazzarella 2023).

Reposting displays two levels of ambiguity. First, there's an ambiguity
about the \emph{meaning} of a repost. As mentioned above, it's unclear
whether reposting conveys endorsement or a weaker attitude (such as
`passing the message along' without endorsing it). Second, there is an
ambiguity about the norms that govern this kind of communicative act.
Offline speech acts, such as promises and assertions, are governed by
norms (e.g. `stick to your promises', `don't lie') that we have
negotiated and refined through generations of cultural evolution. Not so
for reposting: we are still negotiating the norms that govern their use
(Rini 2017), and at least some communicators seem to take those norms to
be very lenient (Altay, de Araujo, and Mercier 2021).\footnote{In
  addition to this, ambiguity is often exacerbated by a phenomenon known
  as context collapse. Broadly, context collapse occurs when distinct
  social spheres or contexts intersect in digital communication, leading
  to difficulties in mutual comprehension (Marwick and boyd 2011;
  McDonald forthcoming). Context collapse can make the interpretation of
  ambiguous acts (such as reposting) more complex, affording the speaker
  even more plausible deniability.} Trading on both of these
ambiguities, reposters can deny any wrongdoing, avoiding some of the
reputational costs associated with spreading misinformation.

There's an additional sense in which reposting comes cheap, in a way
that might contribute to misinformation spread. CST highlights that some
costs of communication are linked to the \emph{production} of a signal,
rather than its social consequences (like the peacock's self-imposed
handicap). Also in this respect, reposting is an affordable signal. In
real-life conversations, spreading a rumour requires cognitive and
material efforts. Explaining how a cabal of Satanist, cannibalistic
paedophiles controls US politics (as the Qanon conspiracy has it) is a
complex task, which requires time and cognitive investment. Reposting
Qanon conspiracies or anti-vax propaganda, by contrast,~takes a single
click; it's instantaneous. Unlike retelling the story yourself, it
requires minimal cognitive and material effort. Precisely because they
are so cheap and quick to produce, reposts can create cascades that can
reach millions of people in a matter of minutes (Vosoughi, Roy, and Aral
2018).

In sum, reposting is `cheap' along two dimensions. In terms of
\emph{production}, the costs of reposting are so low that they are
negligible. \emph{Reputationally}, reposting has reduced costs too,
since it affords the sender plausible deniability (Marsili et al. 2025).
Presumably, reduced costs also explain why reposters aren't too
epistemically careful: users don't always bother to read what they share
(Ward et al. 2023)\footnote{Ward and colleagues don't provide reliable
  empirical evidence for this claim. They misquote a study (Gabielkov et
  al. 2016) which is often misrepresented as claiming that 59\% of users
  repost without reading. But Gabielkov and colleagues make no such
  claim. In personal communication, the authors indicated that unread
  posts actually accounted for only about 15\% of their dataset (cf.
  Marsili 2021). Better indirect evidence for this phenomenon comes from
  studies conducted by Twitter engineers, which have attempted to tackle
  the phenomenon of sharing without reading (Tameez 2020).} and reposts
have been identified as one of the main conduits for fake news
(Vosoughi, Roy, and Aral 2018).

The availability of a cheap signal that can reap many of the benefits of
costlier ones, in turn, inevitably affects the choices that people make
when they communicate online. This creates an environment in which
reposting is often a more alluring option than saying something
yourself.~Why bother investing energy and time into crafting a message,
putting your reputation on the line, when you can simply repost someone
else's? Indeed, users seem to be influenced by these differential
incentives: retweets constitute about half of the content posted on
Twitter (McClain et al. 2021), a number that has been steadily growing
since 2009, when retweets were introduced (Liu, Kliman-Silver, and
Mislove 2014). Reposting is therefore more \emph{prevalent} online than
any comparable signal (such as reporting someone's views) in offline
speech -- making deniable communication, too, more prevalent online than
offline. As signals characterised by lower reputational risks become
increasingly prevalent, it's natural to worry that their pervasive use
might create fertile ground for the spread of misinformation.

However, excessive focus on the epistemic dangers posed by reposts would
be a mistake. The affordability of reposting might have epistemic
benefits, too. If reposting is similar to other forms of cheap and
deniable speech, we should expect reduced costs to be counterbalanced by
\emph{reduced benefits.} After all, reposting someone else's smart tweet
isn't as good for your reputation as crafting one yourself.
Additionally, we might expect reposts to \emph{influence} audiences less
than direct posts. We saw (in §2) that cheaper signals (such as hedged
claims) are generally less persuasive (Price and Stone 2004). If the
same is true online, we should expect reposts to be less convincing than
direct posts, meaning that reposting would not necessarily be a
\emph{persuasive} way of spreading misinformation.\footnote{This
  dovetails nicely with research on fake news uptake, which shows that
  people tend to take what their peers share with a pinch of salt: even
  if falsehoods can easily go viral, people aren't typically persuaded
  by their content (Allcott and Gentzkow 2017; Pfänder and Altay 2023).}
The `affordability' of reposting presumably cuts both ways: it comes
with both benefits (reduced accountability) and costs (reduced
influence). Recent empirical studies tentatively confirm these
hypotheses (Marsili et al. 2025).

Interestingly, while my focus so far has been on reposting, analogous
issues arise for other novel forms of communication, such as
\emph{liking} (McDonald 2021), \emph{tagging} (Scott 2022, chap. 4), or
\emph{subtweeting} (Neufeld and Woodard 2024). Besides being of general
interest, studying how novel speech acts affect reputational dynamics
can help in the process of designing and evaluating practical methods
for containing epistemic pollution on social media. For example, it
suggests that there is promise in design interventions that aim to
increase costs for otherwise cheap speech acts on these platforms.

Two interventions that go in this direction were implemented by Twitter
engineers in 2020, when quote tweets~were made the default for
retweeting (Hatmaker 2020), and when a pop-up window was introduced to
prompt users to read before sharing (when they tried to retweet links
they hadn't opened; see Vincent 2020). These interventions have shown
some initial positive effects (Tameez 2020). From a CST perspective,
they are promising precisely because they contribute to increasing the
costs of~production associated with retweets, making reposting more
laborious (costlier to produce), and dissuading users from reposting
absent-mindedly (by reminding them of the reputational stakes involved).

\subsection{Gamification}\label{gamification}

The model presented so far emphasises the importance of epistemic
incentives in containing misinformation. As mentioned earlier, however,
speakers are motivated by all sorts of non-epistemic incentives. Nguyen
(2021) points out an important way in which social media alter these
motivations, introducing new goals for our communicative exchanges.
Specifically, Twitter (now X) alters our goals by `gamifying'
communicative interactions:

\begin{quote}
Twitter shapes our goals for discourse by making conversation something
like a game. Twitter scores our conversation. And it does so not in
terms of our own particular and rich purposes for communication, but in
terms of its own pre-loaded, painfully thin metrics: Likes, Retweets,
and Follower counts. (410)
\end{quote}

These new communicative goals aren't necessarily aligned with the
`collective pursuit of truth' postulated by our idealised model of
communication. Users quickly internalise Twitter's measurable metrics
(i.e., likes, follower counts, etc.) as their own goals (a process that
Nguyen dubs `value capture'). This can lead them astray from the pursuit
of epistemic goals: chasing likes often requires prioritising popularity
over epistemic scrutiny -- for instance, by sharing an article with a
juicy title regardless of how reliable its source is. Indeed, viral fake
news articles tend to have titles that are just as attention-grabbing as
they are implausible.\footnote{To illustrate, according to Alberto
  Acerbi, the two top performing fake news stories in 2017 were
  `Babysitter transported into hospital after inserting a baby in her
  vagina' and `FBI seizes over 3,000 penises during raid at morgue
  employee's home' (Acerbi 2019, 134): hardly plausible, but effective
  at garnering interest.}

To be sure, the metrics that Twitter gamifies aren't entirely new.
Speakers often pursue non-epistemic goals (like chasing popularity,
influence, social acceptance, or financial gains) that closely resemble
the metrics (e.g. follower counts) that are gamified on social media. We
have long known that non-epistemic incentives can come into tension with
truthfulness -- as demonstrated by the all-too-familiar cases where
politeness is achieved at the expense of sincerity (`Nice haircut!').
What's novel about gamification on social media is the introduction of
powerful non-epistemic incentives (e.g. metrics) \emph{on top of} the
ones that already exist. Crucially, these metrics are built into the
platform and constantly displayed to users, making the pursuit of
non-epistemic goals more salient and attractive on social media.

Recall that, according to CST, a cost is a genuine deterrent only if
it's \emph{costly enough} to counterbalance the opportunity-costs that
the signaller incurs when they forgo their preferred option. Nguyen
(2021) makes a good case that, by leveraging the seductive tactics of
gamification, social media not only increase engagement, but also create
a powerful drive to chase quantifiable metrics over epistemic goals. In
the language of CST, then, gamification threatens to alter the balance
between costs and preferences: if (epistemic) reputational incentives
stay the same, they might not be \emph{strong enough} to counterbalance
the increased preference for pursuing the non-epistemic goals
supercharged by gamified incentives.\footnote{In game-theoretic terms,
  the conjecture is that social media gamification alters the payoff
  structure of a given `communication game', transforming it into a one
  in which cooperation has a smaller basin of attraction.}

Just like Nguyen's work that inspires it, this characterisation of
gamification on social media is primarily conjectural: it is not backed
up by direct empirical evidence. The proposed CST framework thus
highlights some promising avenues for further empirical research, rather
than providing the final word on these matters. Additionally, it
suggests that there is promise in platform design choices that aim to
realign our communicative goals with truthfulness by making the pursuit
of epistemic goals more rewarding. Indeed, some studies show that
rewarding users (e.g. monetarily) leads them to share a significantly
higher proportion of true headlines, and that these interventions have
long-lasting effects (meaning that sharing behaviour is maintained when
the monetary rewards are removed) (Ceylan, Anderson, and Wood 2023). At
least in principle, then, gamification techniques can also be exploited
for the common good, to reshape social incentives in ways that are
truth-conducive: there is space for design choices that realign our
goals with our collective epistemic interests.

\subsection{Ineffective vigilance: the practical challenges of big
data}\label{ineffective-vigilance-the-practical-challenges-of-big-data}

Monitoring is essential for containing the spread of misinformation: if
infractions aren't identified and sanctioned often enough, anybody can
get away with deception. Online communication throws a few spanners in
the works. Let's start by considering a key new challenge, namely the
advent of big data (Floridi 2019, chap. 5). The volume of data
circulating globally has reached staggering heights, with estimates
indicating that we now generate approximately 120 zettabytes of
information yearly -- a number that increased tenfold since 2015 (Edge
Delta 2024), and that was simply inconceivable before the internet. Not
only can we access a disproportionate amount of data, but we can do it
anytime and anywhere. Trips to the library are now obsolete, as
smartphones bring the knowledge of the whole internet to our fingertips.
The result has been an exponential increase in the amount of information
that we access and consume daily.

Since the dawn of the information age, researchers have highlighted how
the availability of vast streams of information is both a blessing and a
curse. We have limited cognitive resources, energy, and time to filter
what's relevant from what's not, and to sort what's true from what's
false. Some scholars suggest that having access to too much information
can actually hinder our ability to learn from it and to make effective
decisions. This phenomenon is known under a variety of buzzwords:
\emph{information overload} (Kovach and Rosenstiel 2010)\emph{,
infostorms} (Hendricks and Hansen 2016), \emph{information glut,} or
\emph{data smog} (Shenk 1998). The idea is that, as the magnitude of
information available increases, our ability to effectively process it
and utilise it diminishes (Bawden and Robinson 2020).

In a similar fashion, as cognitive load increases, our ability to
effectively monitor testimony decreases, potentially making us more
prone to letting our cognitive biases unduly influence which content we
deem true or `shareworthy' (Menczer and Hills 2020). Some studies
suggest that over time, information overload can make individuals more
vulnerable to misinformation (Menczer and Hills 2020; Bermes 2021). More
often, however, people keep their guard too high: unable to monitor all
content, people misclassify \emph{true} content as false more often than
they misclassify \emph{false} content as true (Pfänder and Altay 2023),
with negative effects on information policing and news intake (Rini
2021; Harris 2024a; 2024b, pt. I). From a CST perspective, what matters
is that where information overload hinders epistemic vigilance,
\footnote{Whether being vigilant is actually more difficult in online
  spaces is yet to be established. Digital technologies also offer
  solutions to deal with these challenges. For instance, several social
  media companies employ both AI algorithms and human moderators to
  detect and remove false information, while X (now followed by
  Facebook) introduced Community Notes, where users can add context to
  potentially misleading posts. Despite undeniable limitations, these
  solutions can be helpful (Allen et al. 2021; Pennycook and Rand 2019;
  Wood and Porter 2019; Chuai et al. 2023; Harris 2024, pt. II), meaning
  that some forces \emph{facilitating} vigilance are also at play in
  some online spaces.} the costs associated with sharing misinformation
correspondingly decrease. Information overload, then, may also
contribute to altering the balance of incentives that sustain truthful
communication.

This idea can be articulated more precisely. To establish whether we
should accept a piece of testimony, we rely both on its \emph{content}
(`Is it plausible, i.e., coherent with our prior beliefs?') and its
\emph{source}. Source monitoring, in turn, involves assessing both
\emph{benevolence} (`Is this speaker likely to be sincere?') and
\emph{competence} (`Are they likely to get things right?') (Sperber et
al. 2010). Now, thanks to digital technologies, the practical and
financial costs required for producing and diffusing information have
plummeted. While older media were more centralised networks, with a few
influential nodes broadcasting to many recipients, `networked
{[}digital{]} media allows anyone to be a media outlet' (boyd 2010, 54).
As sources multiply indefinitely, assessing them along both dimensions
(benevolence and competence) becomes more difficult.

Additionally, digital technologies allow misinformation spreaders to
multiply their identities, and shed them when they lose credibility.
Providers of fake news often run `fake news clusters': large-scale
operations composed of many small websites spreading misinformation on
different sub-topics (Papadogiannakis et al. 2023) that appear to work
independently. An example is the network of over 400 fake news websites
operated by Mike Adams (NaturalNews.com, NewsTarget.com, Science.news,
etc.; ISD 2020)\footnote{Of course, readers are often unaware of this
  interdependence. As a result, when a story is amplified by multiple
  outlets from the network, most readers acquire misleading higher-order
  evidence about the credibility of the story (i.e., it looks more
  credible because different outlets published it) -- leading to higher
  levels of persuasion (cf. the literature on the \emph{illusory truth
  effect,} Dechêne et al. 2010)\emph{.} Similar mechanisms can enhance
  the spread of hate speech through reinforcement (Popa-Wyatt 2023,
  795--96).}.

This process of source multiplication poses two related threats to the
social policing of misinformation. First, fake news operators can evade
some of the reputational costs associated with sharing misinformation:
when a site is exposed as fraudulent, dozens of others can still
continue to operate. The same would be more difficult in a non-digital
world, where multiplying news operations would require a
disproportionately larger financial and logistical effort.\footnote{Relatedly,
  one might suspect that digital communication makes epistemic vigilance
  harder because we lack access to non-verbal cues for deception: `By
  barricading our access to these sorts of tells, digital media
  introduce a fundamental asymmetry between the fraudster's capacity to
  mask his lies and his victim's ability to detect deceit' (Reimann
  2022). However, an increasingly large body of research show that such
  non-verbal cues don't actually help us to detect deception (Levine
  2016, chap. 2), meaning that online fraudsters are in no better
  position in this respect.} Second, source monitoring becomes harder as
we become exposed to more sources than we can track the reputation
of.\footnote{Despite this, people are actually quite good at discerning
  the reliability of news sources (Schulz, Fletcher, and Popescu 2020),
  and there is evidence that these skills improve with techniques such
  as `prebunking' and `inoculation' (Compton 2013; Kozyreva,
  Lewandowsky, and Hertwig 2020; van der Linden 2023), within limits
  (Pennycook and Rand 2021, 396; Williams 2023).}

The key takeaway here is that effective vigilance against misinformation
presents new challenges online, due to the volume of content and sources
that require monitoring. An important corollary is that the
truth-conduciveness of different online spaces will vary depending on
their features. Platforms where vast quantities of content are consumed
at an extremely fast pace, such as Instagram and TikTok, likely
exacerbate these difficulties. In contrast, platforms that allow for
slower browsing (and those that implement countermeasures designed to
help us assess the reliability of sources) likely present less severe
challenges.

\subsection{Opaque sources: anonymous, pseudonymous, and artificial
speakers}\label{opaque-sources-anonymous-pseudonymous-and-artificial-speakers}

CST relies on a fairly idealised model of communication, where it's
assumed (i) that receivers can keep track of the communicative record of
other signallers, and (ii) that signallers have an interest in
preserving a good reputation. However, on social media, (i*) it's often
difficult to establish the identity of the signaller, and (ii*)
signallers aren't always motivated to cultivate a good reputation. As we
shall see, both factors may contribute to misinformation spread. To
capture both complications, it will be useful to introduce new
terminology. \emph{Opaque sources}, as I define them,\footnote{The term
  `opacity' is sometimes used to refer to other features of online
  environments, like the inscrutability of recommendation algorithms
  (Miller and Record 2013) or of the inner workings of AI systems
  (Andrada, Clowes, and Smart 2023).} are sources that cannot fully be
held accountable for their communicative behaviour, due to some feature
of digital communication that prevents such accountability, like (i*)
and (ii*).

A paradigmatic case of \emph{opacity} is anonymity. Anonymity challenges
our mechanisms for reputation management in both ways. First, it
undermines our ability to monitor \emph{source reliability}. Anonymous
sources are, by definition, sources whose identity we cannot easily
identify; their reliability cannot be inferred from their reputational
track record. Second, \emph{anonymous} communicators don't have to worry
about staining their reputation: they don't pay the usual costs
associated with sharing false information (like compromising their
credibility, or losing social standing). Therefore, reputational costs
can hardly motivate anonymous speakers to be truthful.

The strategic exploitation of anonymity to avoid reputational costs long
predates the internet era.\footnote{See, for example, discussion of
  anonymity in relation to witchcraft accusations and other costly
  rumours (Favret-Saada 1980; Douglas 2004; Boyer 2018); for a brief
  analysis in relation to hate speech, see Popa-Wyatt 2023, 800.} Rather
than providing a completely new way to escape reputational costs, then,
digital technologies supercharge our ability to do so: online
communication differs in the ease with which anonymity can be achieved,
and in how common it is. Since anonymous speakers don't have to worry
about reputational costs, their increased prevalence, in turn,
facilitates the production and circulation of misinformation online.

With anonymity, however, there is also an interesting trade-off between
reputational costs and influence. All things being equal, costlier
signals should be more reliable than cheaper ones, and hence regarded as
more persuasive (cf. §2). Conversely, anonymous assertions (which are
cheaper) should be less reliable overall, and hence considered less
persuasive by rational agents. This issue has been discussed in
epistemology, too, where diminished reliability has been linked to
diminished epistemic warrant in testimony. Goldberg (2013), for
instance, suggests that anonymous testimony differs from usual testimony
in that we aren't pro tanto entitled to accept it (but cf. Ivy 2021).

If anonymous signals are cheaper but less persuasive, speakers should be
motivated, at times, to forfeit their anonymity to increase credibility.
This is what happens on some platforms. On imageboards like 4chan, users
developed mechanisms to forfeit anonymity, such as `timestamping'
(posting a picture of oneself with the current date and time), unique
Unicode strings , and platform-enabled `tripcodes' (Bernstein et al.
2011; Acerbi 2019). These practices reintroduce means to track source
identity, restoring costly signals despite these platforms' emphasis on
anonymity.

To be sure, genuine anonymity isn't so common online:
\emph{pseudonymity} is more typical (think of Reddit, Discord, or
TikTok). Online pseudonymity enables speakers to hide behind an internet
nickname (Fulda 2007; Véliz 2018; Paterson 2020). Unlike genuine
anonymity, however, pseudonymity doesn't fully evade the costs
associated with communication, nor does it leave an information-seeker
completely clueless about the track record of the communicator. We can
often know whether a pseudonymous source is reliable: for example, on X
(at the time of writing), @SwiftOnSecurity is known to be a reliable
pseudonymous source on cybersecurity advice, whereas @catturd2 is known
for spreading conspiracy theories and disinformation.

Is pseudonymous testimony epistemically sound? On the one hand,
pseudonymous sources can develop a track record within a platform. This
gives receivers some means of assessing their reliability. Additionally,
since pseudonymous users are identifiable in future interactions on the
same platform, they have an incentive to maintain their reputation
within that domain. Pseudonymous testimony, then, should be more
reliable (carrying stronger testimonial warrant) than anonymous
testimony.

On the other hand, pseudonyms allow users to insulate reputational costs
within a specific conversational domain, such as a particular digital
platform. The falsehoods one spreads through a pseudonymous `fake
person' (Harris 2023b) don't necessarily carry over to one's real-life
reputation. This can significantly reduce the epistemic costs attached
to spreading misinformation online, meaning that pseudonymous testimony
is still less reliable (and conveys weaker epistemic warrant) than
standard testimony. Additionally, pseudonymity makes it easy to
misrepresent one's identity, by pretending to be someone or by claiming
properties one doesn't possess. This hinders effective source monitoring
and facilitates misinformation spread, as many ordinary ways of
assessing reliability and honesty (e.g. information about education,
gender, personal motivations, etc.) can easily be fabricated online
(Harris 2022; 2023b; Frost-Arnold 2023, chap. 3).

However, the prevalence of anonymity and pseudonymity need not lead to
unreliable communication: Wikipedia is a prime example. Despite relying
on anonymous and pseudonymous contributions, Wikipedia maintains
reliability through strong community oversight and incentives for
reliable contributions (Fallis 2008; Paternotte and Lageard
2018).\footnote{For a discussion of some additional ways in which
  anonymity can be epistemically beneficial, see Frost-Arnold 2014; 2023
  and Ivy 2021.} Additionally, design interventions such as user
verification can help users discern between pseudonymous and real
sources, improving collective vigilance. When Elon Musk began selling
verification on X, users impersonating companies and politicians began
proliferating, spreading misinformation and causing economic disruption
(Sardarizadeh 2022; B. Y. Lee 2022). This illustrates the importance of
design interventions (such as user verification) that \emph{facilitate},
rather than disrupt, the userbase's ability to assess source reliability
in spite of the widespread presence of opaque sources. Without such
interventions, monitoring opaque sources can become a terribly difficult
task, allowing misinformation to proliferate.

I have focused so far on the opacity of \emph{human} sources.
AI-generated content also bears the hallmarks of opacity. Bots often
masquerade as human communicators (Shao et al. 2018), and cannot be held
accountable for their communicative behaviour in the same way that human
communicators are (Green and Michel 2022; Harris 2023b; Butlin and
Viebahn, forth). While our communicative behaviour is influenced by
considerations about reputation and social influence, bots simply don't
engage in decisions of this sort. Considerations about reputation and
social standing play no direct role (at least currently) in determining
whether their outputs are reliable. So, while concerns about the impact
of generative AI on misinformation are often overblown (Simon, Altay,
and Mercier 2023), it's important to underscore that LLMs also display a
key feature of opacity\footnote{Important caveats apply here. First,
  given that AI agents aren't motivated by reputational concerns, we
  should expect recipients to be wary of their testimony, mitigating
  their potential to serve as a vehicle for misinformation spread.
  Second, reputational concerns presumably affect how these technologies
  are designed -- as proven by efforts to overcome LLM's propensity for
  `hallucinating' (Ji et al. 2023). Many other interesting issues arise
  about artificial agents, but I have no space to address them here;
  I'll take them up in future work instead.} -- they live, in an
important sense, outside of the reputational infrastructure postulated
by CST.

\subsection{\texorpdfstring{Some positives: reach and persistence
}{Some positives: reach and persistence }}\label{some-positives-reach-and-persistence}

In popular culture, online misinformation is often associated with the
democratisation of media. Anybody with a computer and an internet
connection can start a rumour that could potentially be seen by millions
of people all over the world -- as illustrated by the infamous
Macedonian teenagers who, during the 2016 U.S. presidential election,
created over 100 pro-Trump websites spreading fake news (Hughes and
Waismel-Manor 2021), often outperforming mainstream news stories. While
success isn't guaranteed, in offline communication this potential is
typically absent: only a few offline media providers can communicate
with such large audiences. Media scholars use the term \emph{reach}
(Baym 2010) or \emph{scalability} (danah boyd 2010) to highlight this
affordance of online environments.

Against mainstream narratives, however, an increase in reach doesn't
necessarily favour misinformation: there are countervailing forces. As
reach grows, reliable information can get to larger audiences too. A
wider audience means that content goes through more scrutiny, which
increases the chances that falsehoods will be spotted and weeded out
(Acerbi 2019). Finally, as the reach of a message increases, so do its
potential reputational costs. It is one thing to whisper a rumour to a
friend's ears, another to broadcast it to thousands of followers.
Increased reach comes with stronger reputational incentives to be
truthful, too.

Of course, further reach also increases the \emph{rewards} associated
with communicating a message, both in terms of \emph{influence} (it can
persuade much larger audiences) and \emph{reputation} (more people might
acknowledge your authorship). Generalising, reach raises the stakes of
communication: higher rewards are accompanied by higher risks (Reimann
2022, sec. 2.1). While increased costs will be sufficient to put off
some misinformation spreaders, then, others will be willing to take the
risk.

A further complication is that even large audiences can be targeted
quite selectively online, as the internet is composed of many large but
fairly isolated groups (Nguyen 2018). Say you want to spread a wacky
conspiracy (e.g. Pizzagate). Offline, it might be difficult to identify
a community that is ripe for this sort of rumour-mongering (i.e., a
community of dupes that are likely to accept it): finding one would
involve some trial and error, and might impose practical and
reputational costs. Online, finding a ripe audience only takes seconds:
it's as simple as throwing the right keywords into Reddit's search
engine (or into Facebook Groups', or any other platform hosting online
communities).

Targetability can in fact \emph{invert} the payoffs of communicating
falsehoods.\footnote{Sophisticated tools (such as those developed by
  Cambridge Analytica) can help identify targets who are most likely to
  accept a message based on their pre-existing attitudes. While these
  techniques might not be effective at \emph{changing} their target's
  minds (Gibney 2018), they can be of some help when the target is
  already predisposed to accept a message.} Within groups of like-minded
people, the truthfulness of a message is less of a reputational concern
than its conformity with the beliefs of the community. In a sufficiently
uniform group, misinformation spreaders can score positive reputational
points (e.g. by sharing ``proofs'' that the earth is flat within a
community of flat earthers), provided that their message is what the
members are willing to accept. While indiscriminate reach increases the
costs associated with misinformation, then, targeted reach can actually
decrease (or even eliminate) such costs.

Another feature of online communication is that it's often
\emph{persistent} and \emph{searchable} (boyd 2010; Baym 2010). Offline
speech is ephemeral: an embarrassing gaffe or a lie are soon forgotten;
an embarrassing or deceptive tweet is accessible and retrievable for
posterity\footnote{Limitations apply, of course: its author can decide
  to delete it, and one day X's servers will cease to be running.}. And
while offline writing surely enjoys some persistence, it's not
\emph{searchable} like most online content. Searchability increases the
accessibility of any given message: a printed text is only consultable
by those who can find a copy of it; online content can be retrieved by
anyone who uses the right search engine.

\emph{Persistence} and \emph{searchability} can significantly increase
the costs of misinformation. Many social media interactions are
recorded, indexed, and searchable, meaning higher chances that
norm-violations will be detected and punished. The digital footprints we
leave on social media are riskier than we often realise. The tweets that
we absent-mindedly consign to the test of time will be there to haunt us
in the far future.

Take the example of Mark Tykocinski, former president of Thomas
Jefferson University. Tykocinski had to resign from his position after
it was revealed that he liked contentious tweets. Communicative acts
(likes) that he perceived as personal and transient turned out to be
accessible and persistent, making it possible for his controversial
conduct to be detected and punished. \emph{Persistence} and
\emph{searchability} make online communication a risky business. A key
difference with \emph{reach} is that some people are blissfully unaware
of it.

Not all online content is persistent and searchable. Some platforms,
such as Snapchat, owe their popularity precisely to the absence of these
features, which guarantees some degree of privacy and allows the
proliferation of risky (often risqué) messages. Platforms that
prevalently deal with non-textual media, such as Instagram and TikTok,
also display limited degrees of searchability. Content there is
typically more transient: it's consumed for a few seconds, and not
easily searchable, making it hard to find it again. Searchability and
persistence are therefore not a universal feature of online media: they
vary in degree from platform to platform. But where they are present,
they contribute to increased communicative costs and rewards.

Searchability can also be leveraged to make reputation-based
interventions at a platform level. In the wake of the Capitol riots of
2021, Twitter engineers exploited searchability to identify 70,000 users
with a track record of spreading misinformation. After these users were
removed, there was a noticeable reduction in misinformation circulating
through these channels (McCabe et al. 2024)\footnote{The authors
  highlight two important caveats: first, the study cannot fully
  separate the effect of the insurrection from the effects of Twitter's
  policy; second, Twitter only successfully identified a portion of the
  misinformation spreaders. One might have additional concerns about the
  effectiveness of these policies. Some misinformation spreaders can
  migrate to other platforms, such as Trump's `Truth Social'.
  Additionally, moderation policies can fuel concerns about freedom of
  speech; even if unwarranted, such concerns might bolster the
  credibility of misinformation-spreaders among their audiences (Stewart
  2021, 939; Harris 2024b).}. By helping to identify and sanction
misinformation-spreaders, then, searchability can contribute to
increasing the costs of communicating falsehoods.

\subsection{Conclusions}\label{conclusions}

This chapter illustrated how the theoretical tools of costly signalling
theory can be deployed at the intersection of epistemology and
pragmatics, to better understand some of the dynamics that underlie
online communication and misinformation spread. Specifically, we saw how
\emph{novel speech acts, gamification, information overload, source
opacity, reach,} and \emph{persistence} alter the reputational
infrastructure that motivates speakers to communicate truthfully. On top
of identifying factors that might contribute to both fuelling and to
curb misinformation spread, this chapter provided a proof of concept for
how philosophy can tackle these topics from a new perspective, informed
by scientific work on reputational dynamics.

Some of the limitations of this approach, although evident, deserve
explicit acknowledgment. First, there is more to misinformation than CST
alone can address. Many powerful forces fuelling misinformation spread
(such as cognitive biases, or political and economic interests) fall
beyond its remit of inquiry. Second, even within the scope of CST, my
discussion was primarily focused on epistemic incentives: it explored a
few key themes that illustrate the productiveness of the
approach\footnote{For instance, this chapter focused on epistemic
  reputational incentives, but non-epistemic incentives can be equally
  important motivational factors in online communication (Brady,
  Crockett, and Van Bavel 2020). See Bergamaschi Ganapini 2021 and
  Reimann 2022 and for a discussion of some related themes within a CST
  framework.}. Rather than a final, complete picture, then, this chapter
laid out a roadmap for investigating misinformation from a different
perspective. Within these limits, it has demonstrated that CST can help
us better understand the reputational dynamics underlying online
communication. CST thus offers a novel, insightful framework for both
theoretical and empirical research of truthful communication, with
important practical applications for fighting misinformation.

\subsection{References}\label{references}

Acemoglu, Daron, Asuman Ozdaglar, and James Siderius. 2024. `A Model of
Online Misinformation'. \emph{Review of Economic Studies} 91 (6):
3117--50. https://doi.org/10.1093/restud/rdad111.

Acerbi, Alberto. 2019. \emph{Cultural Evolution in the Digital Age}.
Oxford, New York: Oxford University Press.

Acerbi, Alberto, Sacha Altay, and Hugo Mercier. 2022. `Fighting
Misinformation or Fighting for Information?' \emph{Harvard Kennedy
School Misinformation Review}, January.
https://doi.org/10.37016/mr-2020-87.

Adler, Jonathan E. 1997. `Lying, Deceiving, or Falsely Implicating'.
\emph{Journal of Philosophy} 94 (9): 435--52.

Allcott, Hunt, and Matthew Gentzkow. 2017. `Social Media and Fake News
in the 2016 Election'. \emph{Journal of Economic Perspectives} 31 (2):
211--36.

Allen, Jennifer, Antonio A. Arechar, Gordon Pennycook, and David G.
Rand. 2021. `Scaling up Fact-Checking Using the Wisdom of Crowds'.
\emph{Science Advances} 7 (36): eabf4393.
https://doi.org/10.1126/sciadv.abf4393.

Alston, William P. 2000. \emph{Illocutionary Acts and Sentence Meaning}.
Ithaca: Cornell University Press.

Altay, Sacha, de Araujo ,Emma, and Hugo and Mercier. 2022. `\,``If This
Account Is True, It Is Most Enormously Wonderful'':
Interestingness-If-True and the Sharing of True and False News'.
\emph{Digital Journalism} 10 (3): 373--94.
https://doi.org/10.1080/21670811.2021.1941163.

Altay, Sacha, Anne-Sophie Hacquin, and Hugo Mercier. 2020. `Why Do so
Few People Share Fake News? It Hurts Their Reputation.' \emph{New Media
\& Society} 24 (November). https://doi.org/10.1177/1461444820969893.

Altay, Sacha, Rasmus Kleis Nielsen, and Richard Fletcher. 2022.
`Quantifying the ``Infodemic'': People Turned to Trustworthy News
Outlets during the 2020 Coronavirus Pandemic'. \emph{Journal of
Quantitative Description: Digital Media} 2 (August).
https://doi.org/10.51685/jqd.2022.020.

Andrada, Gloria, Robert W. Clowes, and Paul R. Smart. 2023. `Varieties
of Transparency: Exploring Agency within AI Systems'. \emph{AI \&
SOCIETY} 38 (4): 1321--31. https://doi.org/10.1007/s00146-021-01326-6.

Arielli, Emanuele. 2018. `Sharing as Speech Act'. \emph{Versus} XLVII
(2): 243--58. https://doi.org/10.14649/91354.

Baumeister, Roy F., Liqing Zhang, and Kathleen D. Vohs. 2004. `Gossip as
Cultural Learning'. \emph{Review of General Psychology} 8 (2): 111--21.
https://doi.org/10.1037/1089-2680.8.2.111.

Bawden, D., and L. Robinson. 2020. `Information Overload: An Overview'.
In \emph{Oxford Encyclopedia of Political Decision Making}, edited by
David P. Redlawsk. Oxford: Oxford University Press.
https://doi.org/10.1093/acrefore/9780190228637.013.1360.

Baym, Nancy K. 2010. \emph{Personal Connections in the Digital Age}. 1st
edition. Cambridge, UK\,; Malden, MA: Polity Press.

Bergamaschi Ganapini, Marianna. 2021. `The Signaling Function of Sharing
Fake Stories'. \emph{Mind \& Language} 38 (1): 64--80.
https://doi.org/10.1111/mila.12373.

Bermes, Alena. 2021. `Information Overload and Fake News Sharing: A
Transactional Stress Perspective Exploring the Mitigating Role of
Consumers' Resilience during COVID-19'. \emph{Journal of Retailing and
Consumer Services} 61 (July):102555.
https://doi.org/10.1016/j.jretconser.2021.102555.

Bernstein, Michael, Andrés Monroy-Hernández, Drew Harry, Paul André,
Katrina Panovich, and Greg Vargas. 2011. `4chan and /b/: An Analysis of
Anonymity and Ephemerality in a Large Online Community'.
\emph{Proceedings of the International AAAI Conference on Web and Social
Media} 5 (1): 50--57. https://doi.org/10.1609/icwsm.v5i1.14134.

Besnier, Niko. 2019. `Gossip in Ethnographic Perspective'. In \emph{The
Oxford Handbook of Gossip and Reputation}, edited by Francesca Giardini
and Rafael Wittek, 100--118. New York: Oxford University Press.
https://doi.org/10.1093/oxfordhb/9780190494087.013.6.

boyd, Danah. 2010. `Social Network Sites as Networked Publics:
Affordances, Dynamics, and Implications'. In \emph{Networked Self:
Identity, Community, and Culture on Social Network Sites}, edited by
Zizi Papacharissi.

boyd, Danah, Scott Golder, and Gilad Lotan. 2010. `Tweet , Tweet ,
Retweet\,: Conversational Aspects of Retweeting on Twitter'. In
\emph{Proceedings of the 43rd Hawaii International Conference on System
Sciences}, 1--10.

Boyer, Pascal. 2018. \emph{Minds Make Societies: How Cognition Explains
the World Humans Create}. New Haven\,; London: Yale University Press.

Brady, William J., M. J. Crockett, and Jay J. Van Bavel. 2020. `The MAD
Model of Moral Contagion: The Role of Motivation, Attention, and Design
in the Spread of Moralized Content Online'. \emph{Perspectives on
Psychological Science} 15 (4): 978--1010.
https://doi.org/10.1177/1745691620917336.

Brown, Penelope, and Stephen C. Levinson. 1987. \emph{Politeness: Some
Universals in Language Usage}. Cambridge University Press.

Bruner, Justin P. 2024. `Assertions: Deterrent or Handicap? A Reply to
Graham (2020)'. \emph{Episteme}. https://doi.org/10.1017/epi.2023.58.

Budak, Ceren, Brendan Nyhan, David M. Rothschild, Emily Thorson, and
Duncan J. Watts. 2024. `Misunderstanding the Harms of Online
Misinformation'. \emph{Nature} 630 (8015): 45--53.
https://doi.org/10.1038/s41586-024-07417-w.

Butlin, Patrick, and Emanuel Viebahn. 2025. `AI Assertion'. \emph{Ergo}.

Camp, Elisabeth. 2018. `Insinuation, Common Ground, and the
Conversational Record'. In \emph{New Work on Speech Acts}, edited by
Daniel Fogal, Daniel W. Harris, and Matt Moss, 40--66. Oxford: Oxford
University Press.

Ceylan, Gizem, Ian A. Anderson, and Wendy Wood. 2023. `Sharing of
Misinformation Is Habitual, Not Just Lazy or Biased'. \emph{Proceedings
of the National Academy of Sciences} 120 (4): e2216614120.
https://doi.org/10.1073/pnas.2216614120.

Chuai, Yuwei, Haoye Tian, Nicolas Pröllochs, and Gabriele Lenzini. 2023.
`Did the Roll-Out of Community Notes Reduce Engagement With
Misinformation on X/Twitter?' arXiv.
https://doi.org/10.48550/arXiv.2307.07960.

Compton, Josh. 2013. `Inoculation Theory'. In \emph{The SAGE Handbook of
Persuasion: Developments in Theory and Practice, 2nd Ed}, 220--36.
Thousand Oaks, CA, US: Sage Publications, Inc.

Dechêne, Alice, Christoph Stahl, Jochim Hansen, and Michaela Wänke.
2010. `The Truth About the Truth: A Meta-Analytic Review of the Truth
Effect'. \emph{Personality and Social Psychology Review} 14 (2):
238--57. https://doi.org/10.1177/1088868309352251.

Domínguez-Armas, Álvaro, and Andres Soria-Ruiz. 2021. `Provocative
Insinuations'. \emph{Daimon Revista Internacional de Filosofia}, no. 84,
63--80. https://doi.org/10.6018/daimon.481891.

Domínguez-Armas, Álvaro, Andrés Soria-Ruiz, and Marcin Lewiński. 2023.
`Provocative Insinuations as Hate Speech: Argumentative Functions of
Mentioning Ethnicity in Headlines'. \emph{Topoi} 42 (2): 419--31.
https://doi.org/10.1007/s11245-023-09894-6.

Donath, Judith. 2007. `Signals in Social Supernets'. \emph{Journal of
Computer-Mediated Communication} 13 (1): 231--51.
https://doi.org/10.1111/j.1083-6101.2007.00394.x.

Dores Cruz, Terence D., Annika S. Nieper, Martina Testori, Elena
Martinescu, and Bianca Beersma. 2021. `An Integrative Definition and
Framework to Study Gossip'. \emph{Group \& Organization Management} 46
(2): 252--85. https://doi.org/10.1177/1059601121992887.

Douglas, Mary, ed. 2004. \emph{Witchcraft Confessions and Accusations}.
London: Routledge. https://doi.org/10.4324/9780203708545.

Edge Delta. 2024. `Data Market Size: 11 Essential Statistics Unveiling
Trends and Growth'. Edge Delta. 22 March 2024.
https://edgedelta.com/company/blog/data-market-size-and-forecast.

Enders, Adam M., Joseph E. Uscinski, Michelle I. Seelig, Casey A.
Klofstad, Stefan Wuchty, John R. Funchion, Manohar N. Murthi, Kamal
Premaratne, and Justin Stoler. 2023. `The Relationship Between Social
Media Use and Beliefs in Conspiracy Theories and Misinformation'.
\emph{Political Behavior} 45 (2): 781--804.
https://doi.org/10.1007/s11109-021-09734-6.

Fallis, Don. 2008. `Toward an Epistemology of Wikipedia'. \emph{Journal
of the American Society for Information Science and Technology} 59 (10):
1662--74.

Favret-Saada, Jeanne. 1980. \emph{Deadly Words: Witchcraft in the
Bocage}. Cambridge University Press.

Fedeli, Amy Mitchell, Jeffrey Gottfried, Galen Stocking, Mason Walker
and Sophia. 2019. `Many Americans Say Made-Up News Is a Critical Problem
That Needs To Be Fixed'. \emph{Pew Research Center} (blog). 5 June 2019.
https://www.pewresearch.org/journalism/2019/06/05/many-americans-say-made-up-news-is-a-critical-problem-that-needs-to-be-fixed/.

Floridi, Luciano. 2019. \emph{The Logic of Information: A Theory of
Philosophy as Conceptual Design}. Edited by Luciano Floridi. Oxford:
Oxford University Press.
https://doi.org/10.1093/oso/9780198833635.003.0005.

Foster, Eric K. 2004. `Research on Gossip: Taxonomy, Methods, and Future
Directions'. \emph{Review of General Psychology} 8 (2): 78--99.
https://doi.org/10.1037/1089-2680.8.2.78.

Frank, Robert H. 1996. `The Political Economy of Preference
Falsification: Timur Kuran's Private Truths, Public Lies'. \emph{Journal
of Economic Literature} 34 (1): 115--23.

Frigerio, Aldo, and Maria Paola Tenchini. 2023. `Sharing'. \emph{Ratio}
36 (2): 147--58. https://doi.org/10.1111/rati.12375.

Frost-Arnold, Karen. 2014. `Trustworthiness and Truth: The Epistemic
Pitfalls of Internet Accountability'. \emph{Episteme} 11 (1): 63--81.
https://doi.org/10.1017/epi.2013.43.

---------. 2023. \emph{Who Should We Be Online? A Social Epistemology
for the Internet}. 1st ed. New York: Oxford University Press.
https://doi.org/10.1093/oso/9780190089184.003.0001.

Fulda, Joseph. 2007. `The Ethics of Pseudonymous Publication'.
\emph{Journal of Information Ethics} 16 (2): 75--89.

Gabielkov, Maksym, Arthi Ramachandran, Augustin Chaintreau, and Arnaud
Legout. 2016. `Social Clicks: What and Who Gets Read on Twitter?'
\emph{ACM SIGMETRICS Performance Evaluation Review} 44 (1): 179--92.
https://doi.org/10.1145/2964791.2901462.

Gibney, Elizabeth. 2018. `The Scant Science behind Cambridge Analytica's
Controversial Marketing Techniques'. \emph{Nature}, March.
https://doi.org/10.1038/d41586-018-03880-4.

Goldberg, Sanford C. 2011. `Putting the Norm of Assertion to Work: The
Case of Testimony'. In \emph{Assertion: New Philosophical Essays},
edited by Jessica Brown and Herman Cappelen, 175--96. Oxford: Oxford
University Press. https://doi.org/10.1093/acprof.

---------. 2013. `Anonymous Assertions'. \emph{Episteme} 10 (May 2013):
135--51. https://doi.org/10.1017/epi.2013.14.

Goldberg, Sanford C., and David Henderson. 2006. `Monitoring and
Anti-Reductionism in the Epistemology of Testimony'. \emph{Philosophy
and Phenomenological Research} 72 (3): 600--617.

Goldman, Alvin. 2010. `Systems-Oriented Social Epistemology'.
\emph{Oxford Studies in Epistemology} 3:189--214.

Grafen, Alan. 1990. `Biological Signals as Handicaps'. \emph{Journal of
Theoretical Biology} 144 (4): 517--46.
https://doi.org/10.1016/S0022-5193(05)80088-8.

Graham, Peter J. 2020. `Assertions, Handicaps, and Social Norms'.
\emph{Episteme} 17 (3): 349--63. https://doi.org/10.1017/epi.2019.53.

Graham, Peter J., and Nikolaj J. L. L. Pedersen. 2024. `Knowledge Is Not
Our Norm of Assertion'. In \emph{Contemporary Debates in Epistemology},
edited by Blake Roeber, John Turri, Matthias Steup, and Ernest Sosa, 3rd
ed., 339--53. New York: Routledge.

Green, Mitchell. 2009. `Speech Acts, the Handicap Principle and the
Expression of Psychological States'. \emph{Mind \& Language} 24 (2):
139--63. https://doi.org/10.1111/j.1468-0017.2008.01357.x.

---------. 2023. `On the Genealogy and Potential Abuse of Assertoric
Norms'. \emph{Topoi} 42:357--68.
https://doi.org/10.1007/s11245-023-09908-3.

Green, Mitchell, and Jan G. Michel. 2022. `What Might Machines Mean?'
\emph{Minds and Machines} 32 (2): 323--38.
https://doi.org/10.1007/s11023-022-09589-8.

Grice, H. P. 1989. \emph{Studies in the Way of Words}. Cambridge, MA:
Harvard University Press.

Grinberg, Nir, Kenneth Joseph, Lisa Friedland, Briony Swire-Thompson,
and David Lazer. 2019. `Fake News on Twitter during the 2016 U.S.
Presidential Election'. \emph{Science} 363 (6425): 374--78.
https://doi.org/10.1126/science.aau2706.

Grodniewicz, J P. 2022. `Effective Filtering: Language Comprehension and
Testimonial Entitlement'. \emph{The Philosophical Quarterly} 74 (1):
291--311. https://doi.org/10.1093/pq/pqac064.

Harris, Keith Raymond. 2022. `Real Fakes: The Epistemology of Online
Misinformation'. \emph{Philosophy \& Technology} 35 (3): 83.
https://doi.org/10.1007/s13347-022-00581-9.

---------. 2023a. `Beyond Belief: On Disinformation and Manipulation'.
\emph{Erkenntnis}, June. https://doi.org/10.1007/s10670-023-00710-6.

---------. 2023b. `Liars and Trolls and Bots Online: The Problem of Fake
Persons'. \emph{Philosophy \& Technology} 36 (2): 35.
https://doi.org/10.1007/s13347-023-00640-9.

---------. 2024a. `Higher-Order Misinformation'. \emph{Synthese} 204
(4): 127. https://doi.org/10.1007/s11229-024-04763-0.

---------. 2024b. \emph{Misinformation, Content Moderation, and
Epistemology: Protecting Knowledge}. 1st ed. New York: Routledge.
https://doi.org/10.4324/9781032636900.

Hatmaker, Taylor. 2020. `Changing How Retweets Work, Twitter Seeks to
Slow down Election Misinformation \textbar{} TechCrunch'.
\emph{TechCrunch}, 10 September 2020.
https://techcrunch.com/2020/10/09/twitter-retweet-changes-quote-tweet-election-misinformation/?guce\_referrer=aHR0cHM6Ly93d3cuZ29vZ2xlLmNvbS8\&guce\_referrer\_sig=AQAAAD9SWmBNbBqtlIk73Iz82Y9qvU59lc2Sob7yeBiBoaI8om3MWDXCHzIpNEAIdUc9IdmIpbikQ9asnCxL\_s8GXolwmCdf8f5PRRZaYdvsTlQRGfOpKPq9U6h9DVWbomF3p4RWlaOSJWOf-qfbkcsJQJbeMtI5b7Il\_dp47gpu98P-\&guccounter=2.

Hendricks, Vincent F., and Pelle G. Hansen. 2016. \emph{Infostorms: Why
Do We `like'? Explaining Individual Behavior on the Social Net.} Cham:
Springer International Publishing.
https://doi.org/10.1007/978-3-319-32765-5.

Hughes, Heather C., and Israel Waismel-Manor. 2021. `The Macedonian Fake
News Industry and the 2016 US Election'. \emph{PS: Political Science \&
Politics} 54 (1): 19--23. https://doi.org/10.1017/S1049096520000992.

Irons, William. 2001. `Religion as a Hard-to-Fake Sign of Commitment'.
In \emph{Evolution and the Capacity for Commitment: Volume III in the
Russell Sage Foundation Series on Trust}, edited by Randolph Nesse,
290--309. New York, NY, US: Russell Sage Foundation.

ISD. 2020. `Anatomy of a Disinformation Empire: Investigating
NaturalNews'. Institute of Strategic Dialogue.
https://www.isdglobal.org/wp-content/uploads/2020/06/20200620-ISDG-NaturalNews-Briefing-V4.pdf.

Isella, Margherita, Patricia Kanngiesser, and Michael Tomasello. 2019.
`Children's Selective Trust in Promises'. \emph{Child Development} 90
(6): e868--87. https://doi.org/10.1111/cdev.13105.

Ivy, Veronica. 2021. `\textquotesingle Yikkity Yak, Who Said That?' The
Epistemology of Anonymous Assertions'. In \emph{Applied Epistemology},
edited by Jennifer Lackey, 457--81. Oxford University Press.

Ji, Ziwei, Nayeon Lee, Rita Frieske, Tiezheng Yu, Dan Su, Yan Xu, Etsuko
Ishii, Ye Jin Bang, Andrea Madotto, and Pascale Fung. 2023. `Survey of
Hallucination in Natural Language Generation'. \emph{ACM Computing
Surveys} 55 (12): 248:1-248:38. https://doi.org/10.1145/3571730.

Kahan, Dan M., Ellen Peters, Erica Cantrell Dawson, and Paul Slovic.
2017. `Motivated Numeracy and Enlightened Self-Government'.
\emph{Behavioural Public Policy} 1 (1): 54--86.
https://doi.org/10.1017/bpp.2016.2.

Kneer, Markus. 2018. `The Norm of Assertion: Empirical Data'.
\emph{Cognition} 177:165--71.
https://doi.org/10.1016/j.cognition.2018.03.020.

---------. 2021. `Norms of Assertion in the United States, Germany, and
Japan'. \emph{PNAS} 118 (37): 3.

Kneer, Markus, and Neri Marsili. 2025. `The Truth about Assertion and
Retraction: A Review of the Empirical Literature'. In \emph{Lying, Fake
News, and Bullshit}, edited by Alex Wiegmann. Bloomsbury.

Kovach, Bill, and Tom Rosenstiel. 2010. \emph{Blur: How to Know What's
True in the Age of Information Overload}. Bloomsbury Publishing USA.

Kozyreva, Anastasia, Stephan Lewandowsky, and Ralph Hertwig. 2020.
`Citizens Versus the Internet: Confronting Digital Challenges With
Cognitive Tools'. \emph{Psychological Science in the Public Interest} 21
(3): 103--56. https://doi.org/10.1177/1529100620946707.

Krebs, John R., and Richard Dawkins. 1984. `Animal Signals: Mind-Reading
and Manipulation'. In \emph{Behavioural Ecology: An Evolutionary
Approach}, edited by J. R. Krebs and N. B. Davies, 380--402. Blackwell
Scientific.

Lee, Bruce Y. 2022. `Fake Eli Lilly Twitter Account Claims Insulin Is
Free, Stock Falls 4.37\%'. Forbes. 12 November 2022.
https://www.forbes.com/sites/brucelee/2022/11/12/fake-eli-lilly-twitter-account-claims-insulin-is-free-stock-falls-43/.

Lee, James J., and Steven Pinker. 2010. `Rationales for Indirect Speech:
The Theory of the Strategic Speaker'. \emph{Psychological Review} 117
(3): 785--807. https://doi.org/10.1037/a0019688.

Levine, Timothy R. 2016. \emph{Duped. Truth-Default Theory and the
Social Science of Lying and Deception}. University of Alabama Press.

Linden, Sander van der. 2023. \emph{Foolproof: Why We Fall for
Misinformation and How to Build Immunity}. 4th Estate.

Liu, Yabing, C Kliman-Silver, and Alan Mislove. 2014. `The Tweets They
Are A-Changin': Evolution of Twitter Users and Behavior'.
\emph{Proceedings of the International AAAI Conference on Weblogs and
Social Media} 8 (1): 305--14.

Majmundar, Anuja, Jon-Patrick Allem, Tess Boley Cruz, and Jennifer Beth
Unger. 2018. `The Why We Retweet Scale'. \emph{PLOS ONE} 13 (10):
e0206076. https://doi.org/10.1371/journal.pone.0206076.

Marsili, Neri. 2016. `Lying by Promising'. \emph{International Review of
Pragmatics} 8 (2): 271--313. https://doi.org/10.1163/18773109-00802005.

---------. 2018. `Truth and Assertion: Rules versus Aims'.
\emph{Analysis} 78 (4): 638--48. https://doi.org/10.1093/analys/any008.

---------. 2021. `Retweeting: Its Linguistic and Epistemic Value'.
\emph{Synthese} 198:10457--83.
https://doi.org/10.1007/s11229-020-02731-y.

---------. 2023a. `Fictions That Purport to Tell the Truth'. \emph{The
Philosophical Quarterly} 73 (2): 509--31.

---------. 2023b. `Towards a Unified Theory of Illocutionary
Normativity'. In \emph{Sbisà on Speech as Action}, by Laura Caponetto
and Paolo Labinaz, 165--93. Cham: Palgrave Macmillan.

---------. 2024. `Fictions That Don't Tell the Truth'.
\emph{Philosophical Studies} 181 (5): 1025--46.
https://doi.org/10.1007/s11098-024-02098-7.

Marsili, Neri, Rocío Lana-Blond, and Markus Kneer. 2025. `Posting and
Reposting: Investigating Reputation, Trust, and Deniability in Online
Communication'. OSF. https://doi.org/10.31234/osf.io/n5f6d.

Marsili, Neri, and Guido Löhr. 2022. `Saying, Commitment, and the
Lying-Misleading Distinction'. \emph{The Journal of Philosophy} 119
(12): 687--98. https://doi.org/10.5840/jphil20221191243.

Marsili, Neri, and Alex Wiegmann. 2021. `Should I Say That? An
Experimental Investigation of the Norm of Assertion.' \emph{Cognition}
212. https://doi.org/10.1016/j.cognition.2021.104657.

Marwick, Alice E., and Danah boyd. 2011. `I Tweet Honestly, I Tweet
Passionately: Twitter Users, Context Collapse, and the Imagined
Audience'. \emph{New Media \& Society} 13 (1): 114--33.
https://doi.org/10.1177/1461444810365313.

Mascaro, Olivier, and Dan Sperber. 2009. `The Moral, Epistemic, and
Mindreading Components of Children's Vigilance towards Deception'.
\emph{Cognition} 112 (3): 367--80.
https://doi.org/10.1016/j.cognition.2009.05.012.

Mazzarella, Diana. 2023. `\,``I Didn't Mean to Suggest Anything like
That!'': Deniability and Context Reconstruction'. \emph{Mind \&
Language} 38 (1): 218--36. https://doi.org/10.1111/mila.12377.

McCabe, Stefan D., Diogo Ferrari, Jon Green, David M. J. Lazer, and
Kevin M. Esterling. 2024. `Post-January 6th Deplatforming Reduced the
Reach of Misinformation on Twitter'. \emph{Nature} 630 (8015): 132--40.
https://doi.org/10.1038/s41586-024-07524-8.

McClain, Colleen, Regina Widjaya, Gonzalo Rivero, and Aaron Smith. 2021.
`The Behaviors and Attitudes of U.S. Adults on Twitter'. Pew Research
Center.
https://www.pewresearch.org/internet/2021/11/15/the-behaviors-and-attitudes-of-u-s-adults-on-twitter/.

McDonald, Lucy. 2021. `Please Like This Paper'. \emph{Philosophy}, May,
1--24. https://doi.org/10.1017/S0031819121000152.

---------. forthcoming. `Context Collapse Online'. In
\emph{Conversations Online}.

McGowan, Mary Kate. 2019. \emph{Just Words. On Speech and Hidden Harm}.
Oxford University Press.
https://doi.org/10.1001/jama.1968.03140090192022.

McIntyre, Lee. 2018. \emph{Post-Truth}. MIT Press.

Menczer, Filippo, and Thomas Hills. 2020. `Information Overload Helps
Fake News Spread, and Social Media Knows It'. \emph{Scientific
American}, 1 December 2020.
https://www.scientificamerican.com/article/information-overload-helps-fake-news-spread-and-social-media-knows-it/.

Mercieca, Jennifer. 2016. `There's an Insidious Strategy behind Donald
Trump's Retweets'. \emph{The Conversation}, 8 March 2016.
http://theconversation.com/theres-an-insidious-strategy-behind-donald-trumps-retweets-55615.

Mercier, Hugo. 2020. \emph{Not Born Yesterday: The Science of Who We
Trust and What We Believe}. \emph{Not Born Yesterday}. Princeton
University Press. https://doi.org/10.1515/9780691198842.

Metaxas, Panagiotis Takis, E. Mustafaraj, K. Wong, L. Zeng, M. O'Keefe,
and S. Finn. 2015. `What Do Retweets Indicate? Results from User Survey
and Meta-Review of Research'. \emph{Proceedings of the International
Conference on Web and Social Media, ICWSM 2015} 9 (1): 658--61.

Metaxas, Panagiotis Takis, and Twittertrails Research Team TTRT. 2017.
`Retweets Indicate Agreement, Endorsement, Trust: A Meta-Analysis of
Published Twitter Research'. \emph{ArXiv Preprint}.

Miller, Boaz, and Isaac Record. 2013. `\,`Justified Belief In A Digital
Age: On The Epistemic Implications Of Secret Internet Technologies'.
\emph{Episteme} 10 (2): 117--34. https://doi.org/10.1017/epi.2013.11.

Mitchell, Jeffrey Gottfried, Mason Walker and Amy. 2020. `Americans See
Skepticism of News Media as Healthy, Say Public Trust in the Institution
Can Improve'. \emph{Pew Research Center} (blog). 31 August 2020.
https://www.pewresearch.org/journalism/2020/08/31/americans-see-skepticism-of-news-media-as-healthy-say-public-trust-in-the-institution-can-improve/.

Møller, Anders Pape. 1987. `Social Control of Deception among Status
Signalling House Sparrows Passer Domesticus'. \emph{Behavioral Ecology
and Sociobiology} 20 (5): 307--11.

Neufeld, Eleonore, and Elise Woodard. 2024. `On Subtweeting'. In
\emph{Conversations Online}, edited by Patrick Connolly, Sanford C.
Goldberg, and Jennifer M Saul. Oxford University Press.

Newman, Nic, Richard Fletcher, Kirsten Eddy, Craig T Robertson, and
Rasmus Kleis Nielsen. 2023. `Reuters Institute Digital News Report
2023'.

Nguyen, C. Thi. 2018. `Echo Chambers and Epistemic Bubbles'.
\emph{Episteme}, 1--21. https://doi.org/10.1017/epi.2018.32.

---------. 2021. `How Twitter Gamifies Communication'. In \emph{Applied
Epistemology}. Oxford University Press.

Oswald, Steve. 2022. `Insinuation Is Committing'. \emph{Journal of
Pragmatics} 198 (September):158--70.
https://doi.org/10.1016/j.pragma.2022.07.006.

Pagin, Peter, and Neri Marsili. 2021. `Assertion'. In \emph{Stanford
Encyclopedia of Philosophy}, edited by Edward N. Zalta, Winter 2021
edition. https://plato.stanford.edu/archives/win2021/entries/assertion/.

Papadogiannakis, Emmanouil, Panagiotis Papadopoulos, Evangelos P.
Markatos, and Nicolas Kourtellis. 2023. `Who Funds Misinformation? A
Systematic Analysis of the Ad-Related Profit Routines of Fake News
Sites'. In \emph{Proceedings of the ACM Web Conference 2023}, edited by
Ying Ding, Jie Tang, Juan Sequeda, Lora Aroyo, Carlos Castillo, and
Geert-Jan Houbert, 2765--76. Austin TX USA: ACM.
https://doi.org/10.1145/3543507.3583443.

Paternotte, Cédric, and Valentin Lageard. 2018. `Trolls, Bans and
Reverts: Simulating Wikipedia'. \emph{Synthese} 198 (1): 451--70.
https://doi.org/10.1007/s11229-018-02029-0.

Paterson, Grace. 2020. `Sincerely, Anonymous'. \emph{Thought} 9 (3):
167--76. https://doi.org/10.1002/tht3.455.

Pennycook, Gordon, and David G. Rand. 2019. `Fighting Misinformation on
Social Media Using Crowdsourced Judgments of News Source Quality'.
\emph{Proceedings of the National Academy of Sciences} 116 (7):
2521--26. https://doi.org/10.1073/pnas.1806781116.

---------. 2021. `The Psychology of Fake News'. \emph{Trends in
Cognitive Sciences} 25 (5): 388--402.
https://doi.org/10.1016/j.tics.2021.02.007.

Pepp, Jessica, Eliot Michaelson, and Rachel Katharine Sterken. 2019.
`What's New About Fake News?' \emph{Journal of Ethics and Social
Philosophy.} XVI (2): 67--94.

Pfänder, Jan, and Sacha Altay. 2023. \emph{Spotting False News and
Doubting True News: A Meta-Analysis of News Judgments}.
https://doi.org/10.31219/osf.io/n9h4y.

Popa-Wyatt, Mihaela. 2023. `Online Hate: Is Hate an Infectious Disease?
Is Social Media a Promoter?' \emph{Journal of Applied Philosophy} 40
(5): 788--812. https://doi.org/10.1111/japp.12679.

Pozzi, Mélinda, and Diana Mazzarella. 2023. `Speaker Trustworthiness:
Shall Confidence Match Evidence?' \emph{Philosophical Psychology} 37
(1): 102--25. https://doi.org/10.1080/09515089.2023.2193220.

Price, Paul C., and Eric R. Stone. 2004. `Intuitive Evaluation of
Likelihood Judgment Producers: Evidence for a Confidence Heuristic'.
\emph{Journal of Behavioral Decision Making} 17 (1): 39--57.
https://doi.org/10.1002/bdm.460.

Reid, A. Scott, Jinguang Zhang, Grace L. Andreson, and Lauren Keblusek.
2020. `Costly Signaling in Human Communication'. In \emph{The Handbook
of Communication Science and Biology}. Routledge.

Reimann, Ritsaart Willem Peter. 2022. `Costly Displays in a Digital
World: Signalling Trustworthiness on Social Media'. \emph{Social
Epistemology}, 1--18. https://doi.org/10.1080/02691728.2022.2150990.

Reuter, Kevin, and Peter Brössel. 2019. `No Knowledge Required'.
\emph{Episteme} 16 (3): 303--21. https://doi.org/10.1017/epi.2018.10.

Rini, Regina. 2017. `Fake News and Partisan Epistemology'. \emph{Kennedy
Institute of Ethics Journal} 27 (2S): E-43-E-64.
https://doi.org/10.1353/ken.2017.0025.

---------. 2021. `Weaponized Skepticism: An Analysis of Social Media
Deception as Applied Political Epistemology'. In \emph{Political
Epistemology}, edited by Elizabeth Edenberg and Michael Hannon, 31--48.
Oxford University Press.
https://doi.org/10.1093/oso/9780192893338.003.0003.

Robbins, Megan L., and Alexander Karan. 2020. `Who Gossips and How in
Everyday Life?' \emph{Social Psychological and Personality Science} 11
(2): 185--95. https://doi.org/10.1177/1948550619837000.

Roberts, Craige. 2012. `Information Structure in Discourse: Towards an
Integrated Formal Theory of Pragmatics'. \emph{Semantics and Pragmatics}
5 (6): 1--69. https://doi.org/10.3765/sp.5.6.

Sardarizadeh, Shayan. 2022. `Twitter Chaos after Wave of Blue Tick
Impersonations'. \emph{BBC}, 12 November 2022.
https://www.bbc.com/news/technology-63599553.

Saul, Jennifer M. 2012. \emph{Lying, Misleading, and What Is Said: An
Exploration in Philosophy of Language and Ethics}. Oxford University
Press.

Saul, Jennifer M. 2017. `Racial Figleaves, the Shifting Boundaries of
the Permissible, and the Rise of Donald Trump'. \emph{Philosophical
Topics} 45 (2): 97--116.

Saul, Jennifer M. 2018. `Negligent Falsehood , White Ignorance , and
False News'. In \emph{Lying: Language, Knowledge, Ethics, Politics},
edited by Elliot Michaelson and Andreas Stokke. Oxford University Press.

Saul, Jennifer M. 2024. \emph{Dogwhistles and Figleaves: How
Manipulative Language Spreads Racism and Falsehood}. Oxford, New York:
Oxford University Press.

Scheufele, Dietram A., Nicole M. Krause, and Isabelle Freiling. 2021.
`Misinformed about the ``Infodemic?'' Science's Ongoing Struggle with
Misinformation.' \emph{Journal of Applied Research in Memory and
Cognition} 10 (4): 522--26.
https://doi.org/10.1016/j.jarmac.2021.10.009.

Schulz, Anne, Richard Fletcher, and Marina Popescu. 2020. `Are News
Outlets Viewed in the Same Way by Experts and the Public? A Comparison
Across 23 European Countries'. SSRN Scholarly Paper. Rochester, NY.
https://papers.ssrn.com/abstract=3657380.

Scott, Kate. 2021. `The Pragmatics of Rebroadcasting Content on Twitter:
How Is Retweeting Relevant?' \emph{Journal of Pragmatics} 184
(October):52--60. https://doi.org/10.1016/j.pragma.2021.07.022.

---------. 2022. \emph{Pragmatics Online}. London\,; New York.

Searle, John R. 1969. \emph{Speech Acts: An Essay in the Philosophy of
Language}. Cambridge University Press.

Shao, Chengcheng, Giovanni Luca Ciampaglia, Onur Varol, Kai-Cheng Yang,
Alessandro Flammini, and Filippo Menczer. 2018. `The Spread of
Low-Credibility Content by Social Bots.' \emph{Nature Communications} 9
(1): 4787. https://doi.org/10.1038/s41467-018-06930-7.

Shenk, David. 1998. \emph{Data Smog: Surviving the Information Glut
Revised and Updated Edition}. New York: Harper Collins.

Simon, Felix M., Sacha Altay, and Hugo Mercier. 2023. `Misinformation
Reloaded? Fears about the Impact of Generative AI on Misinformation Are
Overblown'. \emph{Harvard Kennedy School Misinformation Review},
October. https://doi.org/10.37016/mr-2020-127.

Simon, Felix M, and Chico Q Camargo. 2023. `Autopsy of a Metaphor: The
Origins, Use and Blind Spots of the ``Infodemic''\,'. \emph{New Media \&
Society} 25 (8): 2219--40. https://doi.org/10.1177/14614448211031908.

Smith, John Maynard, and David Harper. 2003. \emph{Animal Signals}.
Oxford Series in Ecology and Evolution. Oxford, New York: Oxford
University Press.

Sperber, Dan. 2013. `Speakers Are Honest Because Hearers Are Vigilant.
Reply to Kourken Michaelian'. \emph{Episteme} 10 (1): 61--71.
https://doi.org/10.1017/epi.2013.7.

Sperber, Dan, Fabrice Clement, Christophe Heintz, Olivier Mascaro, Hugo
Mercier, Gloria Origgi, and Deirdre Wilson. 2010. `Epistemic Vigilance'.
\emph{Mind and Language} 25 (4): 359--93.
https://doi.org/10.1111/j.1468-0017.2010.01394.x.

Sperber, Dan, and Deirdre Wilson. 1995. \emph{Relevance: Communication
and Cognition}. 2nd ed. Blackwell Publishing.

Stalnaker, Robert C. 1978. `Assertion'. In \emph{Pragmatics}, edited by
Peter Cole. Academic Press.

Stewart, Elizabeth. 2021. `Detecting Fake News: Two Problems for Content
Moderation'. \emph{Philosophy \& Technology} 34 (4): 923--40.
https://doi.org/10.1007/s13347-021-00442-x.

Tameez, Hanaa'. 2020. `Following Successful Experiments, Twitter Will
Prompt All Users to Read the Articles They're about to Retweet'.
\emph{Nieman Lab} (blog). 2020.
https://www.niemanlab.org/2020/09/following-successful-experiments-twitter-will-prompt-all-users-to-read-the-articles-theyre-about-to-retweet/.

Tan, Huibang, Tianxiang Jiang, and Ning Ma. 2023. `Why Do People Gossip?
Reputation Promotes Honest Reputational Information Sharing'.
\emph{British Journal of Social Psychology} 62 (2): 708--24.
https://doi.org/10.1111/bjso.12589.

Tenney, Elizabeth R., Barbara A. Spellman, and Robert J. MacCoun. 2008.
`The Benefits of Knowing What You Know (and What You Don't): How
Calibration Affects Credibility'. \emph{Journal of Experimental Social
Psychology} 44 (5): 1368--75.
https://doi.org/10.1016/j.jesp.2008.04.006.

Tibbetts, Elizabeth A., and James Dale. 2004. `A Socially Enforced
Signal of Quality in a Paper Wasp'. \emph{Nature} 432 (7014): 218--22.
https://doi.org/10.1038/nature02949.

Tibbetts, Elizabeth A., and Amanda Izzo. 2010. `Social Punishment of
Dishonest Signalers Caused by Mismatch between Signal and Behavior'.
\emph{Current Biology} 20 (18): 1637--40.
https://doi.org/10.1016/j.cub.2010.07.042.

Véliz, Carissa. 2018. `Online Masquerade: Redesigning the Internet for
Free Speech Through the Use of Pseudonyms'. \emph{Journal of Applied
Philosophy} 36 (4): 643--58.

Viebahn, Emanuel. 2021. `The Lying/Misleading Distinction: A
Commitment-Based Approach'. \emph{Journal of Philosophy} CXVIII (6).

Vincent, James. 2020. `Twitter Is Bringing Its ``Read before You
Retweet'' Prompt to All Users'. The Verge. 25 September 2020.
https://www.theverge.com/2020/9/25/21455635/twitter-read-before-you-tweet-article-prompt-rolling-out-globally-soon.

Vosoughi, Soroush, Deb Roy, and Sinan Aral. 2018. `The Spread of True
and False News Online'. \emph{Science} 359 (6380): 1146--51.
https://doi.org/10.1126/science.aap9559.

Ward, Adrian F., Jianqing (Frank) Zheng, and Susan M. Broniarczyk. 2023.
`I Share, Therefore I Know? Sharing Online Content - Even without
Reading It - Inflates Subjective Knowledge'. \emph{Journal of Consumer
Psychology} 33 (3): 469--88. https://doi.org/10.1002/jcpy.1321.

Watson, Gary. 2004. `Asserting and Promising'. \emph{Philosophical
Studies} 117 (1): 57--77.

Williams, Daniel. 2022. `Signalling, Commitment, and Strategic
Absurdities'. \emph{Mind \& Language} 37 (5): 1011--29.
https://doi.org/10.1111/mila.12392.

---------. 2023. `The Fake News about Fake News'. \emph{Boston Review},
7 June 2023.
https://www.bostonreview.net/articles/the-fake-news-about-fake-news/.

Williamson, Timothy. 1996. `Knowing and Asserting'. \emph{The
Philosophical Review} 105 (4): 489--523.

Wilson, Deirdre, and Dan Sperber. 2002. `Truthfulness and Relevance'.
\emph{Mind} 25:1--41.

Wood, Thomas, and Ethan Porter. 2019. `The Elusive Backfire Effect: Mass
Attitudes' Steadfast Factual Adherence'. \emph{Political Behavior} 41
(1): 135--63. https://doi.org/10.1007/s11109-018-9443-y.

Yang, Kai-Cheng, Francesco Pierri, Pik-Mai Hui, David Axelrod,
Christopher Torres-Lugo, John Bryden, and Filippo Menczer. 2021. `The
COVID-19 Infodemic: Twitter versus Facebook'. \emph{Big Data \& Society}
8 (1): 20539517211013861. https://doi.org/10.1177/20539517211013861.

Zahavi, A., and A. Zahavi. 1997. \emph{The Handicap Principle: A Missing
Piece of Darwin's Puzzle}. Oxford: Oxford University Press.

Zilinsky, Jan, Yannis Theocharis, Franziska Pradel, Marina Tulin, Claes
de Vreese, Toril Aalberg, Ana Sofía Cardenal, et al. 2024. `Justifying
an Invasion: When Is Disinformation Successful?' \emph{Political
Communication} 41 (6): 965--86.
https://doi.org/10.1080/10584609.2024.2352483.

\subsection{Acknowledgements}\label{acknowledgements}

The author would like to thank Mitchell Green, Patrick Connolly, Mihaela
Popa, and two anonymous referees for their helpful comments on this
chapter. I am especially thankful to the members of the \emph{DPERG}
(\emph{Digital Pragmatics and Epistemology Reading Group)} for the many
insightful discussions during our online meetings and for their valuable
feedback on an earlier draft.

\end{document}